\documentclass[12pt,preprint]{aastex}

\newcommand{\kms}{km s$^{-1}$}

\shorttitle{Molecular gas in G79.29+0.46}
\shortauthors{Rizzo et al.}

\begin{document}


\title{Discovery of warm and dense molecular gas\\ 
surrounding the ring nebula G79.29+0.46}


\author{J.~R.~Rizzo} 
\affil{
Laboratorio de Astrof\'{\i}sica Espacial y F\'{\i}sica Fundamental,
Apartado 78, E-28691 Villanueva de la Ca\~nada, Spain
}
\email{ricardo.rizzo@laeff.inta.es}

\author{F.~M.~ Jim\'enez-Esteban}
\affil{Observatorio Astron\'omico Nacional, Apartado 112,
E-28803 Alcal\'a de Henares, Spain}
\email{f.jimenez-esteban@oan.es}

\and 

\author{E.~Ortiz}
\affil{Universidad San Pablo CEU, Urb.~Montepr\'{\i}ncipe,
E-28668 Boadilla del Monte, Spain}
\email{ortiz.eps@ceu.es}

\begin{abstract}
We present for the first time the detection of mid-$J$ CO line
emission in the outskirts of an evolved massive star, which indicates 
the presence of warm and/or high density molecular gas. Aiming to 
learn about the interplay between evolved massive stars and their 
surroundings, we have carried out CO $J=2\rightarrow1$, $J=3\rightarrow2$, 
and $^{13}$CO $J=2\rightarrow1$ line observations in a 
$4\arcmin\times4\arcmin$ field around the ring nebula 
\object{G79.29+0.46}, which is illuminated by a strong candidate 
to LBV star. The whole field shows extended predominant emission in both
CO and $^{13}$CO $J=2\rightarrow1$ lines, which probably comes from
the large cloud which contains the star forming region
\object{DR\,15}. When this large-scale emission is removed, minor scales 
features become evident, particularly in the CO $J=3\rightarrow2$
line, strikingly coincident with the ring nebula. The high critical
density of CO $J=3\rightarrow2$ (some $10^4$ cm$^{-3}$) gives
additional support for the association with the massive star, since
high density molecular gas has more chances to survive in such a harsh
environment. This structure may have been produced by the cumulative
effect of a strong steady wind in the LBV stage or earlier, which has
compressed the surviving parent molecular cloud. In addition, immersed
within this CO feature, we have also discovered a higher density clump
(at least several $\sim10^5$ cm$^{-3}$), unresolved by the telescope
and probably having a higher kinetic temperature. Towards the clump,
the presence of a jump of $14-16$ \kms\ in the gas velocity may
indicate the existence of a shockfront. This clump may have been
created by at least one mass eruption, $10^3-10^4$ yr ago. Thus, this 
work shows that not all the molecular gas is destroyed during massive 
star evolution, and consequently we are dealing with a new laboratory 
where one can learn about the mass-loss phenomena associated to the 
brief LBV stage.
\end{abstract}


\section{Introduction}

Massive stars are the most influent sources in the dynamical and 
chemical evolution of the galactic interstellar medium (ISM). Although 
relatively few in number, massive stars are the most powerful, the 
fastest in the evolutionary paths and the richest from a chemical 
viewpoint. Mass-loss phenomena (strong and dense winds, violent mass 
ejections) and a very intense UV field provide a significant input to their
surroundings, which heats, ionizes, dissociates, blows out, shocks and 
enriches the ISM as a whole. 

According to current evolutionary models \citep[see, for example,][]{mae94,lan94}, 
the main sequence of massive stars lasts several 10$^6$~yr, while the 
Wolf-Rayet stage takes some 10$^5$~yr. All intermediate stages last 
typically from 10$^3$ to several 10$^4$~yr. Stellar winds, size, effective 
temperature and UV flux vary during each phase by orders of magnitude, 
while the stellar luminosity remains almost constant. When these changes 
occur, a number of different structures in the circumstellar/interstellar
gas should be produced. Numerical simulations \citep{gar96,mar04} predict the 
formation of several shockfronts in the ISM at scales from tenths to tens of pc.
The possibility of mass eruptions at the intermediate and unstable stages may 
also add a significant contribution to the nebular mass \citep{smi06}.

So far, the study of the ISM surrounding massive stars constitutes an excellent 
method to know about the evolutive history of the exciting source, because it 
is expected that different evolutive stages would put their ``fingerprints'' 
in the circumstellar material. Both neutral and ionized gas have been 
widely observed in the last decades, and today we can recognize many
interstellar bubbles associated to the Main Sequence \citep{arn99,cic03} and 
Wolf-Rayet nebulae \citep{mil93,moo00}, among other galactic structures. In 
recent years, the hot dust surrounding massive stars was the subject of several 
near- and mid-IR observational campaigns, and we know about the presence of 
disks, torus and shells around stars at intermediate evolutive stages 
\citep{bec94,oha03}. 

The search for molecular gas in these environments is still limited, possibly due
to the {\it a-priori} hypothesis by which the parent molecular cloud
around an evolved massive star may have been destroyed soon after star formation.
Even so, the detected low-excitation CO \citep{mar99,not02,riz03a} is 
giving indications that the molecular component would represent an important 
contribution to the total mass of gas. Unfortunately, the low-$J$ lines 
observed to date trace densities as low as few $10^3$ cm$^{-3}$, and the 
link with the stars or the circumstellar material can only be based on the 
morphology of the CO structures, and not on any other physical parameter. 

A special and motivating case is \object{NGC\,2359}. To date, this is the 
only Wolf-Rayet nebula where H$_2$ has been detected at 2\,$\mu$m 
\citep{stl98}. Later, the global distribution of CO has been almost 
simultaneously mapped using the Kitt Peak \citep{riz01a} and SEST 
radiotelescopes \citep{cap01}, showing an important amount of molecular
gas surrounding the optical nebula. Follow-up studies have reported the 
detection of complex molecules \citep{riz01b,riz03b} and multiple layers 
of rather dense gas \citep[$\simeq10^4$ cm$^{-3}$;][]{riz03c}. The origin 
of the molecular gas seems related to stages previous to Wolf-Rayet 
(dynamical lifetimes $t_{\rm dyn} \sim 10^4 - 10^5$ yr). 

This significant result encourages the search for molecular gas in the environs 
of evolved massive stars which have not yet become Wolf-Rayet. Of those, the
luminous blue variable (LBV) stars are probably the best candidates, 
due to the ``instantaneous'' effect of their strong winds and spectacular 
mass ejections over the star surroundings \citep{hum94,not97}. The nested 
shells observed in several cases around LBV nebulae span masses from 0.1 
to 20 M$_\odot$, most of which come from short-lived eruptions, such 
as the well known 19th century outburst of $\eta$\,Car \citep{smi06}.
The multiple layer structure discovered in \object{NGC\,2359}, referred
above, may have been formed during a previous LBV stage.

A recent census provides a total of 12 confirmed galactic LBV stars and 23 
candidates of such objects \citep{cla05}. \object{G79.29+0.46} appears 
as a strong LBV candidate in the sample. Noticed for the first time by 
\citet{wen91} in a radio map (1420\,MHz) of the Cygnus X region, this 
ringlike structure looks highly symmetric, and has a diameter of $\sim 3\farcm 2$.
In a pioneer paper based on continuum and {\sc Hi} observations, \citet{hig94} 
have shown the thermal nature of the continuum emission, and suggested that 
the ringlike nebula is an ionized shell of swept-up interstellar material; 
in addition, a possible {\sc Hi} bubble is reported at a systemic velocity 
(LSR) of $\sim-3$ \kms. By assuming a normal gas to dust ratio, the 
atomic mass is not enough to account for the extinction measured, and 
hence {\bf a significant amount of molecular gas is predicted}. After examining 
the IRAS High Resolution images, \citet{wat96} concluded that 
\object{G79.29+0.46} is a detached shell due to an epoch of high mass loss 
($\sim 5\,10^{-4}$\,M$_{\odot}$\,yr$^{-1}$), followed by a more quiet period. 
They have estimated a shell mass of $\sim$14\,M$_{\odot}$, and measured a dust 
temperature of 60\,--\,77\,K. This picture is consistent with the LBV 
nature of \object{G79.29+0.46}'s exciting star, where some different events of 
mass-loss may have occurred in the recent past.

The central star of \object{G79.29+0.46} has been spectroscopically studied 
at the optical and the near-infrared wavelengths \citep{wat96, voo00}. Its LBV 
classification is supported by a high luminosity ($>$\,10$^{5}$\,L$_{\odot}$) 
and a currently high mass-loss rate ($>10^{-6}$\,M$_{\odot}$ yr$^{-1}$). Its 
galactic longitude is not adequate to estimate a reliable 
kinematic distance; even so, \citet{voo00} derived a value of 
$\approx$\,2\,kpc. This is close to 1.7 kpc, the estimated distance to 
\object{Cygnus OB2} \citep{mas91}, a huge OB association which \object{G79.29+0.46} 
may belong to. \object{G79.29+0.46} is behind the Great Cygnus Rift, projected 
$9\arcmin$ to the north-west of the star-forming region DR15, which causes high 
visual extinction \citep[up to 16\,mag;][]{kno00}, and where an important amount 
of diffuse CO were reported \citep[][and references therein]{sch06}.

The aim of this work is to detect and to make a first morphological and 
dynamical analysis of the molecular gas associated to \object{G79.29+0.46}. 
Below we present the results of surveys in the rotational lines of CO 
$J=2\rightarrow1$ and $3\rightarrow2$, and $^{13}$CO $J=2\rightarrow1$ 
surrounding the LBV ring nebula. The obtained results encourage to make further 
studies about dynamics, excitation and the possible relationship between the 
molecular gas detected and the ring nebula.

\section{Observations}

The observations were made with the Submillimeter Telescope (SMT), located on 
Mount Graham, Arizona. The telescope is a 10.4 m diameter primary with a nutating 
secondary. Design, optics and structural aspects of the telescope have been 
extensively described by \citet{baa99}. The single-polarization receiver SIS-230
was used to observe the CO and $^{13}$CO $J=2\rightarrow1$ line emission; on the
other hand, the CO $J=3\rightarrow2$ line was observed using the dual-channel
SIS-345 receiver. The three lines were mapped in several 
$\sim 4\arcmin\times4\arcmin$ fields around \object{G79.29+0.46}, using 
{\it on-the-fly} mode, with a fixed position (empty of line emission) as the
reference. Several {\it on-the-fly} maps were later averaged in order to get a
final {\it rms} as low as possible. The scanning directions and starting positions
were different each time to reduce stripping. The Table 1 summarizes some parameters
of the observations; for each line, the table indicates the frequency, angular 
resolution (HPBW), main-beam efficiency and final {\it rms} of the maps.

The receivers were tuned for double-sideband operation, assuming a sideband ratio of
1 for the system temperature calibration. The receivers were calibrated with 
ambient-temperature and liquid nitrogen-cooled absorbing loads to determine the 
receiver noise temperature; calibration of the final data have uncertainties below 20\,\%. 
These measurements were made each time a receiver was tuned and at intervals of a few 
hours thereafter. System temperatures were measured with the standard sky/ambient 
temperature load method \citep{kut81}. Calibrations were performed every 15-20 minutes 
to track variations in atmospheric opacity. Typical single-sideband system 
temperatures during the observations were in the range $250-400$\,K for the 1.3\,mm lines 
and $1200-2000$\,K for the 0.8 mm line, depending on the elevation and atmospheric opacity.
Two acousto-optic spectrometers (AOS) have been used as backends. The AOS have 2048 
channels, and a frequency resolution of 480 kHz, providing a velocity resolution of 
0.6 (0.4) \kms\ in the 1.3\,mm (0.8\,mm) bands, and bandwidths wide enough for 
galactic emission. The pointing was checked within 2\,hr on a planet or other bright 
continuum source using the integrated line intensity. Corrections to the expected 
positions for the pointing model were not more than $4\arcsec$ in either azimuth or 
elevation. Planet observations were also used to check the main beam efficiency at the 
observed frequencies, which are also listed in Table 1. Throughout all the paper, the 
temperatures are on a main-beam scale ($T_{\rm MB}$), and the velocities refer 
to the Local Standard of Rest system ($V_{\rm LSR}$).

\section{Results: emission maps around G79.29+0.46}

Figure 1 shows a panel of the emission maps corresponding to the CO $J=2\rightarrow1$
line. To provide a first-look morphological comparison with G79.29+0.46, the circle 
in the center of each map roughly traces the external border of the ring nebula
\citep{hig94}. At first sight, we note that large-scale emission dominates all the 
observed field, mostly at the 2.4 \kms\ map. Despite of this, some minor-scale 
features can be observed, such as the unresolved arc at the southwest, around 
($-80", -110"$). This arc is present in a wide range of velocities, and bounds part
of the ring nebula. The inner part of the circle is filled by some low-level emission, 
especially in the range -3 to +5 \kms. The emission at the first and last maps do not 
appear morphologically linked to the ring nebula.

Figure 2 sketches the $^{13}$CO $J=2\rightarrow1$ emission in a similar fashion to
Fig.~1. With the aim of reducing the map noise, a 2-D hanning filtering has been
applied to build the figure. This filtering slightly reduces the angular resolution
of the map down to $45"$. In general, the emission of this isotopomer is well correlated
with the most intense CO emission in Fig.~1. Similarly, besides the most intense 
emission at the southeastern corner of the observed field, minor scale feature are 
also noticed. A systematic deviation of the contours is visible toward the north, 
north-east, and south-west of the plotted circle. Another outstanding feature of 
Fig.~2 is the lack of emission in the first, second, and last maps; as this 
behaviour can not be explained merely by the sensitivity of the map, it clearly 
indicates that the CO opacity is relatively higher at the central velocities (i.e., 
in the range $-3$ to $+10$ \kms), where the isotopic ratio [CO]/[$^{13}$CO] is 
lower, when compared with the more extreme velocities plotted in Fig.~2.

The emission maps of the CO $J=3\rightarrow2$ line are shown in Fig.~3. Due to a higher
velocity resolution, the twelve maps of the figure span the same velocity range as
those of Figs.~1 and 2. With the exception of the last map (11.4 \kms), the emission 
concentrates on the central velocities, mainly on the six central maps. Some extended
emission appears at -0.5 and 1.2 \kms, while in the following three maps the CO
$J=3\rightarrow2$ emission is concentrated outside the ring nebula. Especially 
outstanding is again the southwest feature; it has the highest level of the maps 
(above 10\,K on a $T_{\rm MB}$ scale), remains unresolved at 345 GHz, and is widely 
extended in velocity.

\section{Linking the CO to G79.29+0.46: distance, morphology, kinematics}

The field of \object{G79.29+0.46} is immersed into Cyg X, a major 
star-forming region which contains large amounts of molecular gas having 
densities of some $10^3$\,cm$^{-3}$ \citep{sch06}. More specifically, 
\object{G79.29+0.46} is projected close in the sky (only 9\arcmin)
to the smaller star-forming region \object{DR\,15}, although it is not clear
if they are physically connected. \object{DR\,15} is located at
$1.7\pm0.3$\,kpc \citep{sch06}. The distance to \object{G79.29+0.46}, 
however, is hard to obtain with similar precision due to the uncertainties 
in the stellar parameters, dependent on its classification. In this sense, 
\citet{voo00} has made an important contribution, providing a distance in 
the range 1 to 4\,kpc, with a likely value of 2\,kpc, close to the distance 
of \object{DR\,15}.

Thus, the CO emission presumably linked to \object{G79.29+0.46} are 
expected to be heavily contaminated by foreground/background emission 
from \object{DR\,15} or \object{Cyg\,X} as a whole. Moreover, 
contribution from the solar neighbourhood does not have to be 
disregarded (mostly in the $J=2\rightarrow1$ lines), due to the 
velocities involved, close to cero.
As we have seen in Figs.~1, 2, and 3, most of the CO
emission is dominated by large-scale emission which fills all the
mapped area. From a visual inspection to the low resolution maps of
\citet{oka01}, we can relate our large-scale CO emission 
--which increases towards the southeast-- with the molecular cloud
around \object{\object{DR\,15}}. Furthermore, in the wide field MSX 
and SCUBA images of \citet{red03} we can clearly associate the CO 
emission to the higher column densities of dust. In principle, we 
can therefore think about the large-scale emission detected in our maps 
as arising from the foreground/background emission associated to the 
molecular cloud where \object{DR\,15} is embedded.

It is also expected that the CO gas associated to the ring nebula 
should have minor-scale features, and be morphologically connected to 
the ring nebula.  Thus, in order to propose a link between the 
observed CO and \object{G79.29+0.46}, we should look at the smaller-scale 
features. In this sense, the southwestern clump is the most prominent
feature, visible in all the maps. Other minor scale features can
be inferred by the local deviations of the contours already referred
to in Sect.~3. 
In order to better distinguish the local features presumably
associated to the ring nebula, we have removed a plane emission 
(i.e., a polynomial of order 1 in both celestial coordinates) for
each velocity channel, in the three observed lines. These planes have been
fitted by a least-square method to the whole field, taking
the external zone of the field as input, and assuming that every
feature outside $\approx 160\arcsec$ is purely associated to the
foreground/background emission. The results are presented in Fig.~4,
where integrated velocity maps in two different velocity ranges ($+1.3$
to $+4.8$ \kms\ at panels 4a to 4c; $-3.0$ to $-1.3$ \kms\ at panels 4d to 4f)
are shown for each line. Both the positive and negative velocity
ranges have been selected to better show the minor-scale features of
the CO emission. The molecular emission is shown by contours,
superposed on an ISO 25\,$\mu$m-image in grayscale \citep{tra97}.

The distribution of the background-substracted emission is
now closely following the ring-shaped geometry of the infrared nebula. 
At the positive velocity range (Figs.~4a to 4c) the most prominent 
feature is the southwest clump, clearly visible in both CO lines. The CO 
$J=3\rightarrow2$ line emission also has lower level contour emission 
bordering the external part of the nebula. The lack of a $^{13}$CO
counterpart of the southwestern clump may be due to the background
removal method, because the intensity of the local feature may be
comparable to the fluctuations of the fitted large-scale plane.
Strikingly, at negative velocities (Figs.~4d to 4f), the emission of 
the three lines closely match the infrared nebula, especially in 
the western half of the ring. The peak of the emission is again at 
the southwest, but at an inner position with respect to the 
southwestern clump at positive velocities.

The morphological coincidence of the small-scale features of the 
CO emission with the infrared ring nebula encourages their mutual 
association. At the best of our knowledge, the CO $J=1\rightarrow0$ and
$J=2\rightarrow1$ observations presented in \object{AG\,Car} is the only
LBV nebula where a molecular counterpart has been reported \citep{not02}. 
In that paper, however, no maps were constructed and it is hard to
determine if the morphology of the CO resembles or not that found in 
our data. Moreover, the low-$J$ lines may contain important amounts of 
diffuse, background or foreground gas. In \object{G79.29+0.46}, 
the small-scale features that more clearly are correlated to the ring 
nebula are those of the $J=3\rightarrow2$ line (Fig.~4c and 4f), and 
they are also the most intense. If we also take into account the higher 
critical density of the $J=3\rightarrow2$ line (some $10^4$ cm$^{-3}$), 
compared to the low-$J$ lines, we can think about a scenario where high 
density molecular gas has more chances to survive in harsh environments, 
dominated by strong and variable stellar winds and UV fields from the 
massive progenitor star.

The kinematics of the local gas was studied by position-velocity maps across selected 
strips. On many of them, we could not detect any sign of expansion or rotation of the 
CO gas. The dominant velocities across the field are from $-2$ to 0 \kms, with almost no 
systematic variations or gradients. This is, on the other hand, the velocity range
where most of the low-resolution emission was reported as arising from \object{DR\,15}
\citep{oka01}. An outstanding exception to this general trend is shown in Fig.~5. This 
figure was constructed by selecting the strip indicated in Fig.~4c, along a northeast to 
southwest direction (exactly 45\degr\ from north to east). The slice was selected to 
pass through the southwest arc (at $\sim 260\arcsec$ in Fig.~5), where 
a sudden broadening occurs. This broadening grows up to 14 \kms\ (from $-2$ to $+12$ 
\kms), and is limited to an angular extension comparable to the beamsize. It is also 
noticeable that the CO velocity shifts when the strip passes from the inner to the 
outer region of the nebula, i.e., when passes through the southwest arc.

The feature displayed in Fig.~5 is a dynamical indicator of a low-velocity
shock, acting beyond the ionized ring nebula at a projected velocity of 14 \kms. 
This assumption is supported by two facts: firstly, the region affected by the 
broadening is relatively thin, because it is unresolved by our beam; and secondly,
there is no smooth variations of the velocity between the inner and the outer 
regions. An important, additional argument is provided by the fact that there is 
no CO emission arising from \object{DR\,15} or other close objects, above 10 \kms;
in this sense, the Fig.~4c of \citet{oka01} is ellocuent, showing no features at 
velocities above 10 \kms. In other words, the CO velocity outside the ring nebula 
is locally exceptional, and it may only be explained by some kind of interaction 
with the LBV-candidate star. The presence of shocked gas close to the hot star 
encourages its association and explains its own existence, because the (relatively) 
recent mass eruptions and variable winds after the main sequence may have produced 
some sort of features like the ones recognized in the molecular circumstellar gas. 

As were mentioned in the introduction, an increasing number of hydrodynamical 
simulations are predicting the presence of several shocks in the environs of 
evolved massive stars \citep{gar96,mar04}. This shocked region is certainly of 
particular interest, because it may constitute a good laboratory to test these 
theoretical predictions. This is especially true when we take into account the
multiple layers reported by \citet{riz03c} surrounding a Wolf-Rayet nebula 
(\object{NGC\,2359}), where a previous LBV stage was assigned as the responsible 
of such layers. The chemistry of the molecular gas in the southwestern clump
may also be of particular interest, with a scenario characterized by a shocked 
cloud, and close to a highly-obscured, hot, and evolved star.

The velocity range of the unshocked CO (from $-3$ to $+1$ \kms) is 
--within the uncertainties of the observations and the expected dispersion-- 
coincident with the {\sc Hi} hole's systemic velocity reported by \citet{hig94}.
This {\sc Hi} counterpart may represent the main sequence signature of the 
star progenitor, as in many other cases of the Galaxy 
\citep[see, for example,][]{arn99,riz01a,cic03}. These velocities do not differ
too much to those of the metal lines \citep{voo00} and Br$\alpha$ 
\citep{wat96} from the star
\footnote{Note that, in both papers, the conversion from 
heliocentric to LSR reference systems were made in the wrong sense, and 
consequently the star and nebular velocities result to be $\sim 34$ \kms\ higher 
than the published ones.}.

Below we discuss the particular physical conditions of the 
molecular gas linked to \object{G79.29+0.46}, and also provide 
some quantitave arguments about its possible origin.

\section{Discussion}
\subsection{Physical parameters: the LVG approach}

As we mentioned in the earlier sub-section, the detected molecular gas should 
have physical conditions which are different to those of the general ISM. The 
detected CO may simply be interstellar gas accumulated by the stellar wind, 
the results of several ejecta, or the relics of the parent molecular cloud, 
affected by the destructive action of UV radiation 
and/or pushing wind. An estimate of the molecular mass, together with the 
neutral atomic and the ionized gas masses, may complete the energetic 
balance related to the mechanical input from the exciting star. Other local 
parameters, such as the gas temperature or the density, will help to better 
describe the physical conditions in the studied region.

In a wide variety of galactic scenarios, CO was found to be close to LTE but 
having a high opacity \citep{leq05}. From the most common values of the CO to 
$^{13}$CO $J=2\rightarrow1$ line ratio (between 3 and 5 when $^{13}$CO 
is detected) we confirm that both CO lines are optically thick. Furthermore, the high 
intensity of the CO $J=3\rightarrow2$ line (comparable or even higher than the 
$J=2\rightarrow1$ line), indicates that we are dealing with densities 
($n$(H$_2$)) above $10^4$ cm$^{-3}$, and rotational temperatures ($T_{\rm rot}$) 
of 40\,K or more.

We can make a first approach to the physical conditions of the CO by using a 
LVG code to predict the observed intensities and line ratios. In order to 
compare the LVG modelling with the observations, we have used the line-integrated 
maps already shown in Fig.~4 in the two velocity intervals. The
CO $J=3\rightarrow2$ map has been convolved to the $J=2\rightarrow1$ angular 
resolution to work with a directly comparable $T_{\rm MB}$ scale. The CO to 
$^{13}$CO $J=2\rightarrow1$ line ratio varies from 2 to 8, while the CO 
$J=3\rightarrow2$ to $2\rightarrow1$ line ratio (hereafter referred to as 
$R_{\rm exc}$) varies from 0.3 to 1.7 on a $T_{\rm MB}$ scale in both velocity
intervals. 

After analyzing the emission in all the lines, as well as the line ratios, we 
have proceeded with the LVG modelling by choosing characteristic regions 
within the observed field. These regions are indicated in Fig.~6, and the 
meanings are as follows:

\noindent {\it Region A:} It contains the most intense emission from the 
southwestern clump in the (1.3, 4.8) \kms\ velocity interval. This region 
may have representative values for both the total mass and $n$(H$_2$) of 
the emission at positive velocities.

\noindent {\it Region B:} This region corresponds to the maximum $R_{\rm exc}$ 
in both velocity intervals, and points out the maximum densities expected in 
the whole area.

\noindent {\it Region C:} This region comprises most of the emission at 
negative velocities (excluding region $B$), coincident with the ring nebula. 
It may represent the most common physical conditions in the velocity interval 
(-3, -1.3) \kms, and may contain most of the mass.

\noindent {\it Region D:} This is a control region, with no significant 
features associated to the ring nebula. This is built just to compare the 
results of LVG modelling with the other regions, and to check the validity 
of the method.

The four regions defined are sketched in Fig.~6 superposed to the distribution 
of $R_{\rm exc}$ in the (1.3, 4.8) \kms\ velocity interval in greyscale, and 
to some contours of the CO $J=3\rightarrow2$ emission in both velocity 
intervals (see caption). Note that greyscale peaks at region B and not 
at region A, where the line intensity is higher for positive velocities.
Qualitatively, region B might have a higher density but lower mass than the 
region A, where we have found the CO absolute peak and a high (but not the 
highest) $R_{\rm exc}$. 

We have modelled the CO emission for different values of the kinetic temperature 
($T_{\rm k}$), namely 40, 70, and 100\,K. For a given $T_{\rm k}$, we have used 
the $^{13}$CO $J=2\rightarrow1$ line intensity and $R_{\rm exc}$ as input, and 
thus predicted the $^{13}$CO column density ($N$($^{13}$CO)) and $n$(H$_2$). 
According to the method, $N$($^{13}$CO) is most sensitive to the line intensity, 
and does not vary too much with $T_{\rm k}$ \citep{leq05}. The obtained values 
for $N$($^{13}$CO) and $n$(H$_2$) were later introduced to obtain the line 
opacities and excitation temperatures. The CO column density ($N$(CO)) was 
obtained from $N$($^{13}$CO) by assuming a CO to $^{13}$CO relative abundance
of 70 \citep{wil92}. Finally, the total molecular mass was derived 
from $N$(CO), by assuming a CO abundance of 8 10$^{-5}$, a distance of 
1.7 kpc, and including the correction for the presence of He and other elements.
The results from the LVG modelling are summarized in Table 2. For each 
velocity interval, we have indicated the region, its angular size, the assumed 
$T_{\rm k}$, and the obtained values for $n$(H$_2$), $N$(CO), and the total 
molecular mass. Finally, the last column provided a brief description of each 
region. 

The visual absorption $A_{\mathrm V}$ measured towards \object{G79.29+0.46} 
varies from 12 \citep{tra99} to 15 mag \citep{voo00}. By assuming a crude 
gas-to-dust ratio of 
$$ N({\rm HI}) + 2\, N({\rm H}_2) \sim 10^{21} A_{\mathrm V}$$
\noindent we derive $N({\rm H}_2) \le 6 - 7.5\ 10^{21}$ cm$^{-2}$. For a normal
CO abundance (8 10$^{-5}$), a general $N$(CO) value of $\le 5 - 6\ 10^{17}$ 
cm$^{-2}$ is expected. The values of $N$(CO) obtained in our control regions 
are one order of magnitude lower, which may indicate that most of the local
gas is in atomic phase (both neutral and ionized). Moreover, the densities of 
the control regions are the lowest of the table, which again reinforces
the proposed link of the molecular gas characterized by the regions A, B, and C.
While the densities of region D are around $10^3$ cm$^{-3}$, the values for 
the other regions show and enhancement of one or two orders of magnitude.

The total mass at positive velocities rounds 8\,M$_\odot$, roughly distributed 
between clumps A (5\,M$_\odot$) and B (3\,M$_\odot$). At negative velocities, a 
total mass of 5\,M$_\odot$ is computed, distributed between the clump B 
(1.5\,M$_\odot$) and the northwestern arc (3.5\,M$_\odot$). The missing mass 
of the method, mostly due to the reduction of the velocity intervals and to the 
selected regions, is in any case lower than 2\,M$_\odot$. These masses fall 
within the range of those expected in LBV outbursts \citep{smi06}, and to other 
verified mass eruptions like in \object{$\eta$\,Carina} \citep{smi03}, or 
\object{P\,Cyg} \citep{mea01}. Although the lack of symmetry of the associated 
CO is puzzling, the hypothesis of mass ejections as the origin of such 
structures is worth to be considered in the future. A follow-up study should 
be carried out to observe other emission lines at higher frequencies or larger 
telescopes, capable of resolving the southwestern structure. The dominant 
physical processes may also be inferred by getting information from molecular 
tracers of shocks and PDRs.

The assumed $T_{\mathrm k}$ is surely not uniform along the whole field: probably, 
the emission at negative velocities is better described by higher $T_{\mathrm k}$ 
than the regions emitting at positive velocities.
As expected, the column densities are almost independent from $T_{\mathrm k}$, and 
the variations are well below the observational uncertainties. 
The $^{13}$CO $J=2\rightarrow1$ line was found to be optically thin in all cases.
However, both CO lines are optically thick; the $J = 3\rightarrow2$ line systematically
has a larger opacity than the $J=2\rightarrow1$ line, which may result in larger values 
of $R_{\rm exc}$, and consequently $n$(H$_2$). 
The $^{13}$CO $J=2\rightarrow1$ line is subthermally excited in all cases. For 
positive velocities, the LVG modelling needs an excitation temperature above 40\,K, 
indicating a $T_{\mathrm k}$ well above such value. In fact, fitting was impossible 
in region B for $T_{\mathrm k}=40$\,K.

\subsection{On the origin of the detected molecular gas}

The high densities found in the field, particularly in the regions A and B, strengthen
the hypothesis of several shockfronts acting upon the circumstellar medium around
\object{G79.29+0.46}. A set of shockfronts is the natural consequence of the 
presence of different wind regimes during star evolution, and the probable occurence 
of mass eruptions \citep{gar96,mar04}. So far, it is important to associate the 
different features with specific evolutive stages of the star progenitor. 

We can deal with a rough estimate of the age of the high-density gas, by taking
into account the morphology and kinematics presented above. For an assumed distance
of 1.7 kpc, the southwestern clump is projected at a distance of $\sim$ 1 pc. A lower
limit for the shock velocity would be 14 \kms, the magnitude of the ``jump'' in Fig.~5.
By adopting these values, a crude estimate of the dynamical time would be
$$t_{\rm dyn} < \frac{R}{V_{\rm shock}} = \frac{1\ {\rm pc}}{14\ {\rm km\,s}^{-1}} 
\simeq 7\times 10^4\ {\rm yr}.$$
This value of $t_{\rm dyn}$ should be regarded as a loose upper limit for the age of the
shock, mainly due to the fact that: ({\it i}) the present shock velocity may be higher 
than 14 \kms, because this is the most conservative value, and just a radial projection 
of the real shock velocity; ({\it ii}) the shock might have started with values of 
several hundreds of \kms, and then have been decelerated when travelling into the 
circumstellar gas. If this were the case, the real age would be one order of magnitude 
lower. By looking at the high regularity of the torus-like nebula, a reasonable value 
for the age of the structure may be below $10^4$ yr, perhaps some $10^3$ yr.

The youth of this structure allows us to associate the molecular 
shockfront with the actual LBV phase, which might have experienced several episodes
of mass ejection in the near past. As the material is indeed more disperse 
than, for example, \object{$\eta$ Carina}, we can speculate about two 
possibilities: ({\it i}) \object{G79.29+0.46} is older than \object{$\eta$ Carina},
and hence has had more time to disperse its surroundings; or ({\it ii}) the
energy involved in \object{G79.29+0.46} is greater than in \object{$\eta$ Carina}.
The first hypothesis seems more plausible, because of the size of the circumstellar
material associated to \object{G79.29+0.46}, and because recent, violent 
eruptions of \object{$\eta$ Carina} have been recorded in the 19th century.
Even so, it is interesting to explore both possibilities, as well as to
go deep into a comparison of both objects, in order to shed some light on 
the interplay between these type of stars and their outskirts.

\section{Conclusions}

A region of $4\arcmin\times4\arcmin$ in the field of the ring nebula
\object{G79.29+0.46} has been mapped in the CO $J=2\rightarrow1$ and
$3\rightarrow2$, and the $^{13}$CO $J=2\rightarrow1$ emission lines,
with a maximum angular resolution of 24\arcsec, equivalent to $\sim
0.2$ pc at an assumed distance of 1.7 kpc. We report for the first
time the detection of a mid-$J$ CO emission line from the surroundings
of a LBV nebula. Provided the excitation conditions of the CO 
$J=3\rightarrow2$ line, the studied gas is presumably warm (more than 
33\,K) and dense (above some $10^4$ cm$^{-3}$). As a first result
of this work, it is shown that not all the molecular gas is destroyed 
during massive star evolution, and consequently we are dealing with a 
new laboratory where one can learn about the mass-loss phenomena 
associated to the brief LBV stage.

The CO emission is dominated by large scale features, which uniformly 
increases towards the southeast; this background/foreground emission 
is probably associated to \object{DR\,15}, due to its close location 
and the velocity range of emission. Minor scale features, however, show 
a morphology completely different, dominated by a large arc which 
coincides with more than half of the infrared ring nebula, and 
bounding it at a projected distance of up to 1pc. This 
emission is particularly intense (in the $J=3\rightarrow2$ line) 
towards the southwest, where a more conspicuous zone is embedded. 
Moreover, there is a shockfront of a projected velocity of 14 \kms\ 
acting precisely in this southwestern clump.

We have predicted the observed line emission by LVG modelling. A minimum
$T_{\rm k}$ of 40\,K was determined in an extended area which have densities 
of the order of $10^4$ cm$^{-3}$. Immersed within this plateau, there are 
two outstanding clumps (regions A and B), which have densities well above 
$10^5$ cm$^{-3}$ and probably a higher $T_{\rm k}$.

The total molecular mass associated to the nebula is $\sim$ 13 to 
15\,M$_\odot$, while the clump masses are about 5 \,M$_\odot$ each. Most 
of the gas may result from the surviving parent molecular cloud, later 
compressed by steady stellar winds and radiation pressure. The origin of 
the clumps may be different, and we can not roule out the existence of 
recent mass eruptions. The higher density, the presence of a shockfront 
and the mass involved may be indicators of violent events, produced 
$\sim 10^4$\,yr ago, or less. This fact, together with the high symmetry 
of the nebula, points to an LBV origin of the feature.

This study of the molecular component directly associated to the LBV stage 
may help to understand the recent past of \object{G79.29+0.46}, and therefore the 
evolutive mechanisms of this kind of objects. So far, it is worth making more 
observational efforts in order to resolve the emission, particulary in the clump. 
The differences between regions A and B are puzzling, and may indicate
the existence of more than one mass eruption in the recent past of the progenitor 
of \object{G79.29+0.46}. A follow-up study of complex molecules should also be
carried out, because they can provide information about the physical parameters 
in detail. The chemistry of the molecular gas in this region may also be of 
particular interest, with a scenario characterized by a shocked cloud, and close 
to a highly-obscured, hot, and evolved star. An abundance study of such 
molecules may point to the dominant emission mechanisms in the zone, which 
can be related to the ionized and atomic neutral gas, together with the hot 
circumstellar dust. Finally, other surveys of CO emission in LBVs and other 
evolved massive stars will also help to understand how frequent the mass 
eruptions are in this class of objects, and how important they are when 
compared to the steady high mass loss and to the UV destructive radiation.

\acknowledgments

We are grateful to the technical staff of SMT for their kind assistance during 
both the remote and on-site observations. We also wish to acknowledge the 
anonymous referee by his/her useful comments, which greatly improved the 
paper. Data reduction and analysis were done using the GILDAS package 
(http://www.iram.fr/IRAMFR/GILDAS).
This work is supported by the Spanish Ministerio de Ciencia y T\'ecnica, 
through grant AYA2003-06473. The SMT is operated by the Arizona Radio Observatory 
(ARO), Steward Observatory, University of Arizona.

Facilities: \facility{ARO(SMT)}.

\clearpage

%
%

\clearpage

\begin{deluxetable}{rrrrr}
\tablecaption{Observational parameters\label{tab1}}
\tablewidth{0pt}
\tablehead{
\colhead{line} & \colhead{frequency} & \colhead{HPBW} & \colhead{$\eta_{\rm MB}$} & 
\colhead{$rms$\tablenotemark{a}} \\
\colhead{}     & \colhead{(GHz)}       & \colhead{($\arcsec$)} & \colhead{}       & 
\colhead{(K)} 
}
\startdata
CO $J=2\rightarrow1$ & 230.5380000 & 36 & 0.85 & 0.4 \\
$^{13}$CO $J=2\rightarrow1$ & 220.3986765 & 38 & 0.86 & 0.2 \\
CO $J=3\rightarrow2$ & 345.7959899 & 24 & 0.60 & 0.6 \\
\enddata


\tablenotetext{a}{Approximate $rms$ of the final maps, in $T_{\mathrm R}^*$ scale}

\end{deluxetable}

\clearpage

\begin{deluxetable}{crrrrrl}
\tablecaption{Physical parameters in selected regions\label{tab2}}
\tablewidth{0pt}
\tablehead{
\colhead{Region} & \colhead{Area} & \colhead{$T_{\rm k}$} & \colhead{$n$(H$_2$)} & \colhead{$N$(CO)}    & 
\colhead{Mass}  & Description\\
\colhead{}       & \colhead{(ster)} & \colhead{(K)}   & \colhead{(cm$^{-3}$)} & \colhead{(cm$^{-2}$)}  & 
\colhead{(M$_\odot$)} &
}
\startdata
\multicolumn{7}{c}{Positive velocities: from 1.3 to 4.8 \kms:}\\
\\
A     & 5.5 $10^{-8}$ & 40 & 2.4 $10^{5}$ & 1.4 $10^{17}$ & 5.2 
      & Southwestern clump. \\
      & &  70 & 1.7 $10^{4}$ & 1.2 $10^{17}$ & 4.5 
      & CO $J=3\rightarrow2$ peak. \\
      & & 100 & 9.1 $10^{3}$ & 1.2 $10^{17}$ & 4.5 \\
\\
B     & 3.9 $10^{-8}$ & 40 & $>4\ 10^{6}$ & 8.3 $10^{16}$ & 2.7 
      & $R_{\rm exc}$ peak. \\
      & &  70 & 1.3 $10^{5}$ & 9.5 $10^{16}$ & 3.1 \\
      & & 100 & 2.0 $10^{4}$ & 8.2 $10^{16}$ & 2.7 \\
\\
D     & 7.6 $10^{-8}$ & 40 & 7.7 $10^{3}$ & 5.1 $10^{16}$ & 2.5 
      & ``Control'' region.\\
      & &  70 & 4.1 $10^{3}$ & 5.0 $10^{16}$ & 2.5 \\
      & & 100 & 2.5 $10^{3}$ & 4.9 $10^{16}$ & 2.4 \\
\hline
\\
\multicolumn{7}{c}{Negative velocities: from -3.0 to -1.3 \kms:}\\
\\
B     & 3.9 $10^{-8}$ & 40 & 2.7 $10^{4}$ & 4.7 $10^{16}$ & 1.6 
      & CO $J=3\rightarrow2$ and \\
      & &  70 & 8.8 $10^{3}$ & 4.5 $10^{16}$ & 1.5 
      & $R_{\rm exc}$ peaks. \\
      & & 100 & 6.4 $10^{3}$ & 4.6 $10^{16}$ & 1.5 \\
\\
C     & 4.0 $10^{-7}$ & 40 & 1.7 $10^{4}$ & 1.3 $10^{16}$ & 3.4 
      & Most extended \\
      & &  70 & 6.5 $10^{3}$ & 1.3 $10^{16}$ & 3.4 
      & emission, coincident\\
      & & 100 & 4.8 $10^{3}$ & 1.3 $10^{16}$ & 3.5 
      & with ring nebula.\\
\\
D     & 7.6 $10^{-8}$ & 40 & 6.4 $10^{3}$ & 2.7 $10^{16}$ & 1.3 
      & ``Control'' region. \\
      & &  70 & 3.2 $10^{3}$ & 2.6 $10^{16}$ & 1.2 \\
      & & 100 & 2.2 $10^{3}$ & 2.6 $10^{16}$ & 1.2 \\
\\
\enddata


\end{deluxetable}

\clearpage


%
%

\begin{figure}
\includegraphics[width=\columnwidth,height=!]{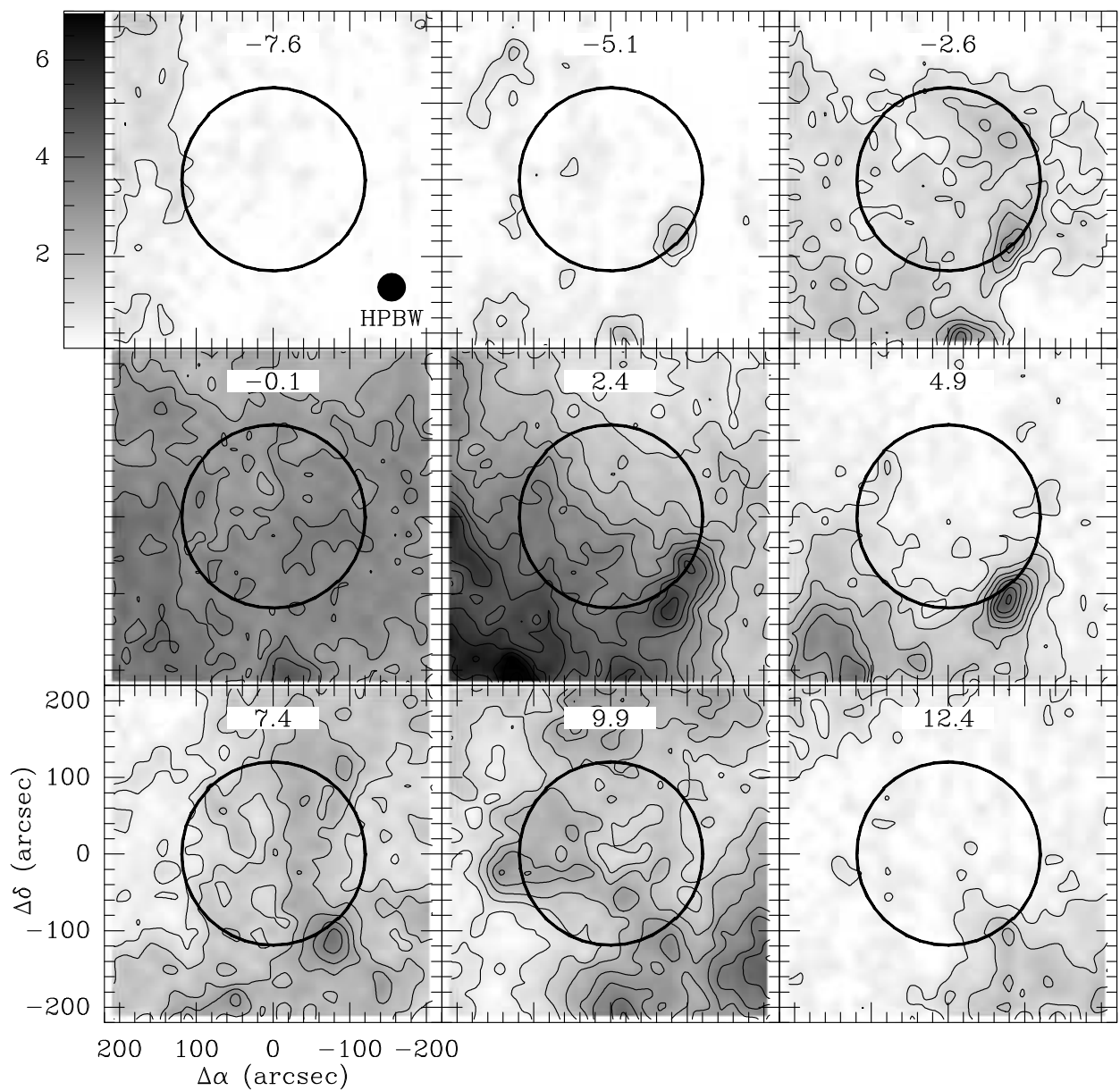}
\caption{CO $J=2\rightarrow1$ emission maps in the field of G79.29+0.46. 
Central velocities (in \kms) are indicated on top of each map, and 
correspond to the mean of two consecutive channels. Starting and spacing 
contours are 0.6\,K (in $T_{\rm MB}$ scale). Greyscale (in K) and beam size 
are indicated in the top left map. The central circle, $\sim4'$ in diameter, 
roughly sketches the external border of the ring nebula. Equatorial 
coordinates (J2000.0) are offsets from the exciting star of G79.29+0.46
\label{fig1}}
\end{figure}

\clearpage

%
%

\begin{figure}
\includegraphics[width=\columnwidth,height=!]{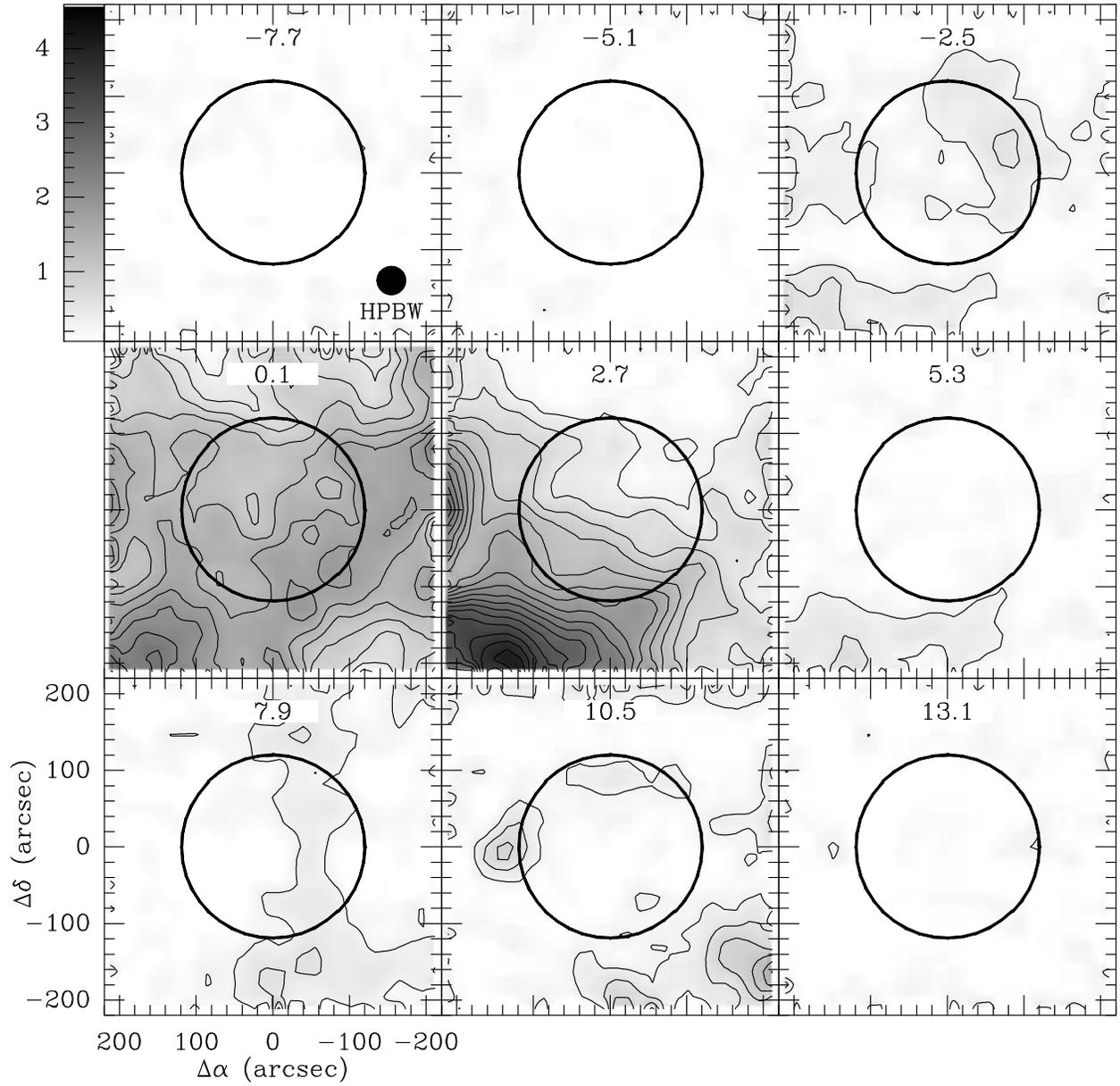}
\caption{
The same as for Fig.~1, but for the $^{13}$CO $J=2\rightarrow1$ line.
Starting and spacing contours are 0.25\,K
\label{fig2}}
\end{figure}

\clearpage

%
%

\begin{figure}
\includegraphics[width=\columnwidth,height=!]{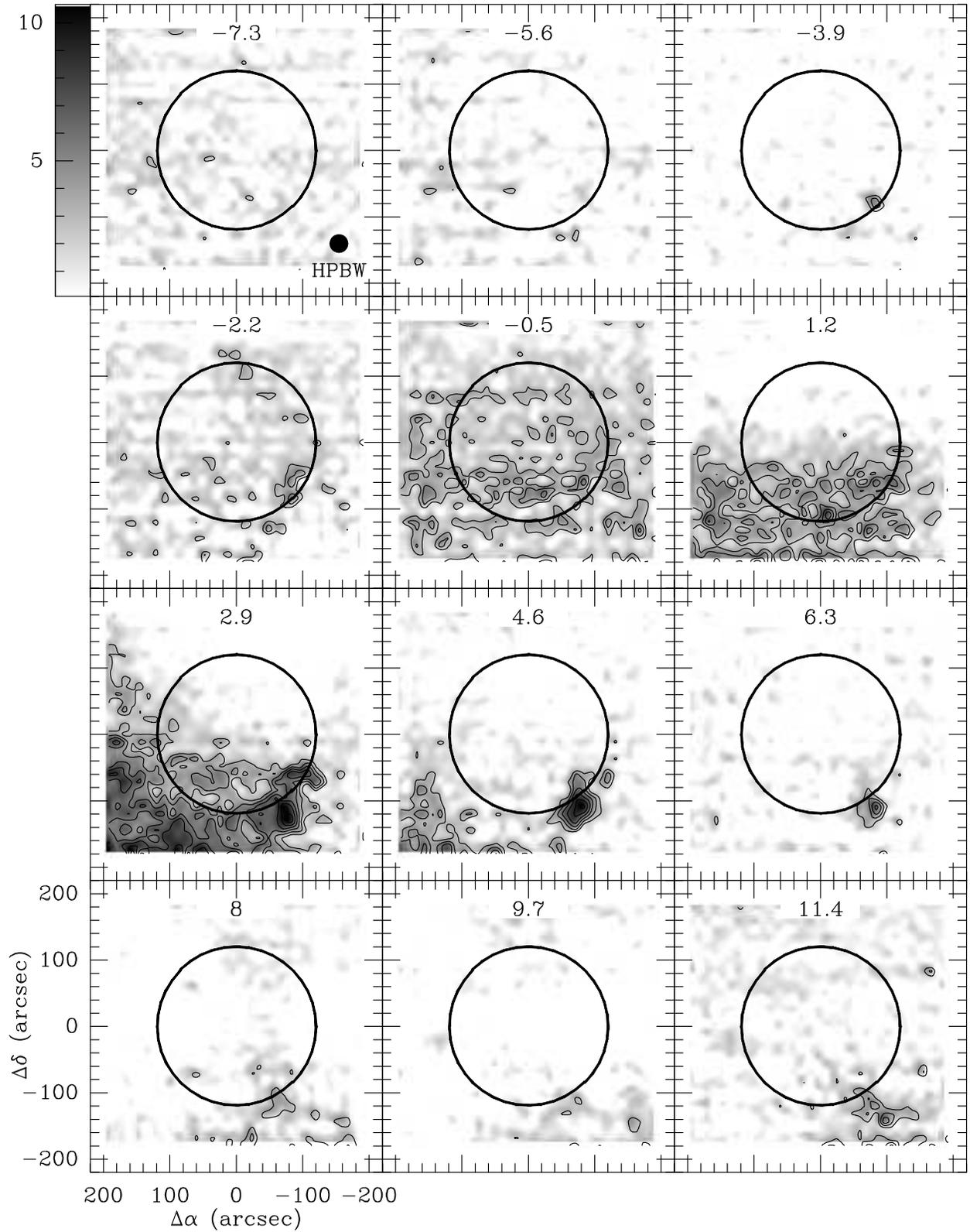}
\caption{
The same as for Figs.~1 and 2, but for the CO $J=3\rightarrow2$ line.
Contours are spaced by 1.5\,K, starting in 3\,K 
\label{fig3}}
\end{figure}

%
%
\begin{figure}[ht]
\includegraphics[width=\columnwidth,height=!]{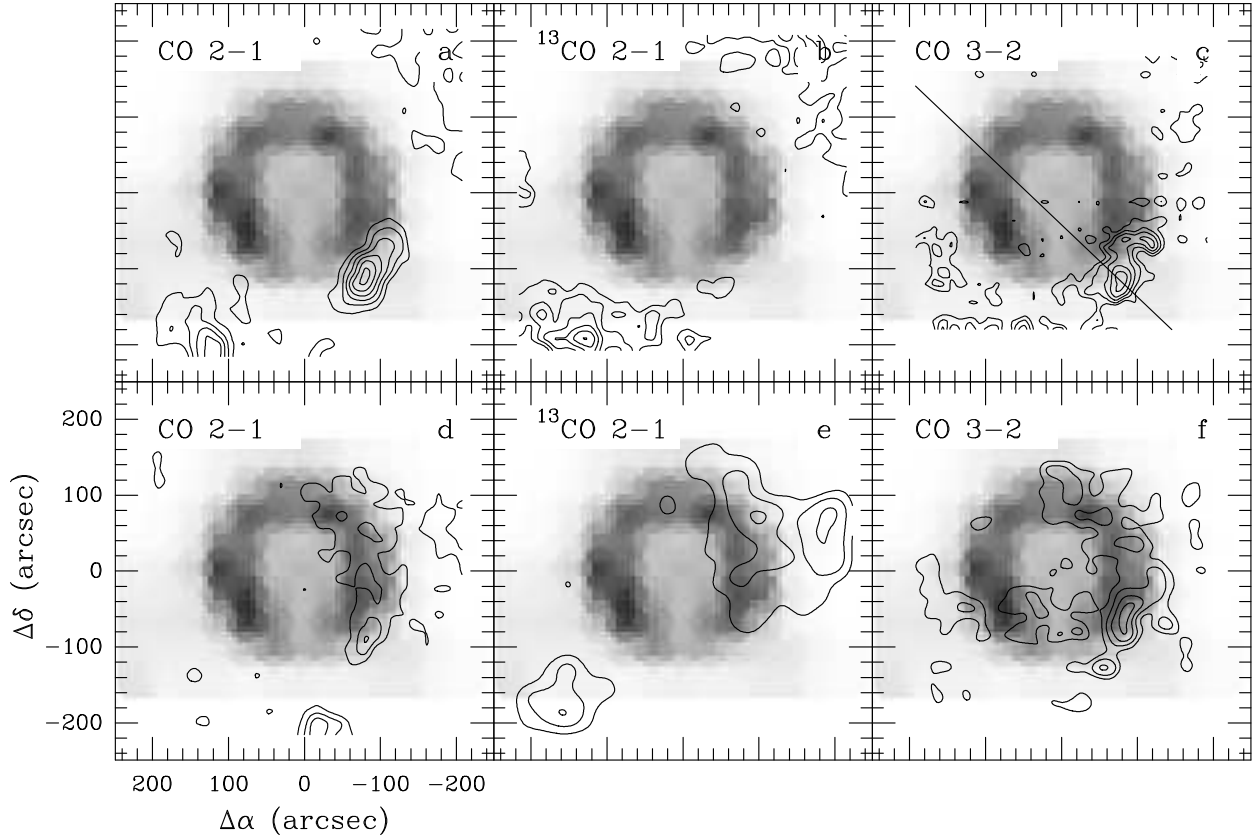}
\caption{Background-removed maps of the three lines observed, integrated in
two different velocity ranges: from 1.3 to 4.8 \kms\ (panels {\bf a} to {\bf c}), and
from $-3.0$ to $-1.3$ \kms\ (panels {\bf d} to {\bf f}). Lines are indicated at the
top left of each map. Contours are overlaid to an ISO 25\,$\mu$m image of 
the nebula \citep{tra97}. Contours start and spacing are 1.6, 1.0, 4.4,
1.0, 0.4, and 1.0 K\,\kms\ from {\bf a} to {\bf f}, respectively. 
The diagonal line on map {\bf c} marks the strip selected for the 
position-velocity diagram shown in Fig.~5
\label{fig4}}
\end{figure}

%
%
\begin{figure}[ht]
\includegraphics[angle=-90,width=14cm,height=!]{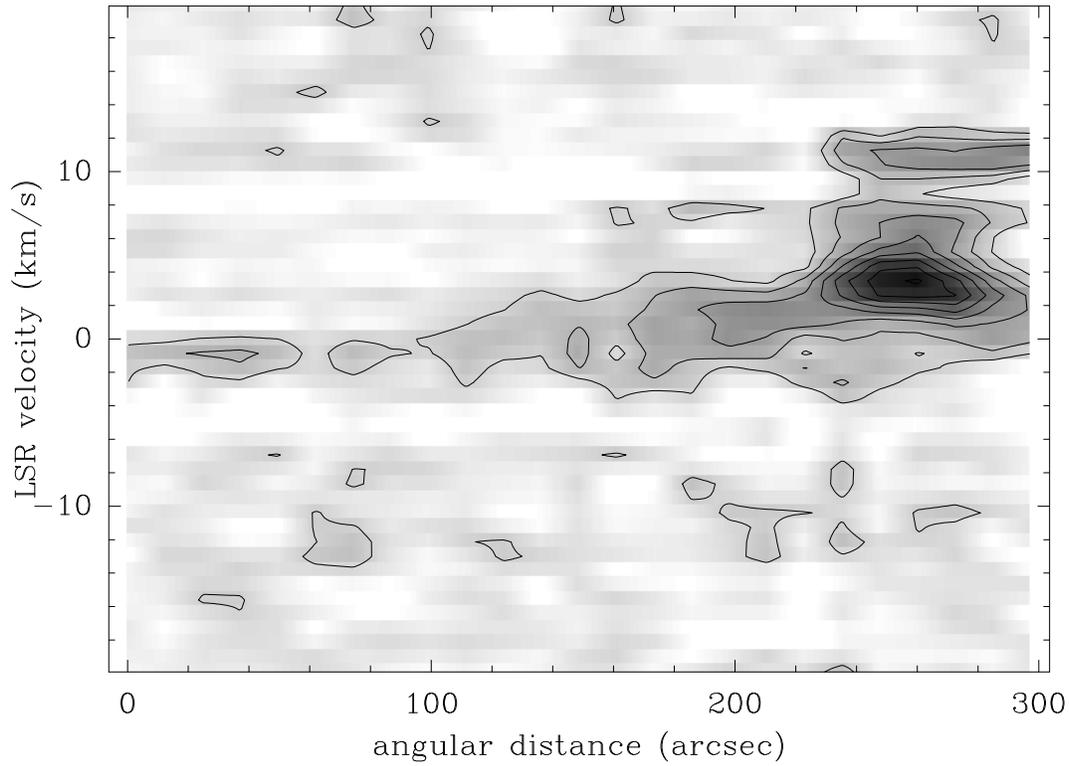}
\caption{Position-velocity diagram of the CO $J=3\rightarrow 2$ line emission in
the field of G79.29+0.46. The direction of the slice is traced by the 
line in Fig.~4c, oriented 45\degr\ from northeast to southwest. The notable 
broadening near 260\arcsec\ corresponds to the southwest arc just beyond the 
nebula (see text)
\label{fig5}}
\end{figure}

%
%
\begin{figure}[ht]
\includegraphics[width=9cm,height=!]{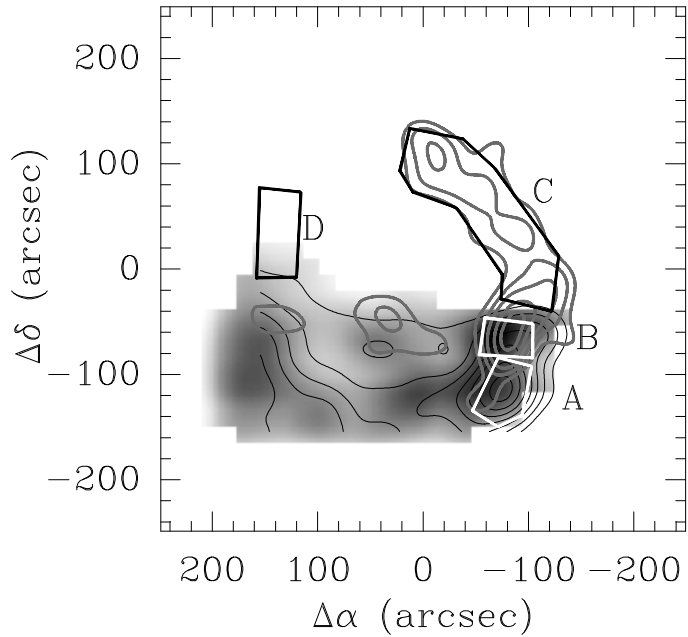}
\caption{Regions selected for the LVG modelling, indicated by letters A, B, C and D.
Greyscale corresponds to the CO $J=3\rightarrow2$ to $2\rightarrow1$ line ratio 
($R_{\rm exc}$) in the velocity interval (+1.3, +4.8) \kms. Black, thin contours 
correspond to the CO $J=3\rightarrow2$ line emission in the same velocity interval.
Grey, heavy contours trace the same but in the velocity interval (-3, -1.3) \kms. 
All data have been convolved to 36\arcsec\ of angular resolution
\label{fig6}}
\end{figure}


\clearpage


\begin{thebibliography}{}

\bibitem[Arnal et al.(1999)]{arn99} Arnal, E.~M., Cappa, 
C.~E., Rizzo, J.~R., \& Cichowolski, S.\ 1999, \aj, 118, 1798 

\bibitem[Baars et al.(1999)]{baa99} Baars, J.~W.~M., Martin, R.~N., 
Mangum, J.~G., McMullin, J.~P., \& Peters, W.~L.\ 1999, \pasp, 111, 627

\bibitem[Becklin et al.(1994)]{bec94} Becklin, E.~E., 
Zuckerman, B., McLean, I.~S., \& Geballe, T.~R.\ 1994, \apj, 430, 774 

\bibitem[Cappa et al.(2001)]{cap01} Cappa, C.~E., Rubio, M., 
\& Goss, W.~M.\ 2001, \aj, 121, 2664 

\bibitem[Cichowolski et al.(2003)]{cic03} Cichowolski, S., 
Arnal, E.~M., Cappa, C.~E., Pineault, S., \& St-Louis, N.\ 2003, \mnras, 
343, 47 

\bibitem[Clark et al.(2005)]{cla05} Clark, J.~S., Larionov, 
V.~M., \& Arkharov, A.\ 2005, \aap, 435, 239 

\bibitem[Garc\'{\i}a-Segura et al.(1996)]{gar96} Garc\'{\i}a-Segura, 
G., Mac Low, M.-M., Langer, N. 1996, A\&A, 305, 229 

\bibitem[Higgs et al.(1994)]{hig94} Higgs, L.~A., Wendker, 
H.~J., \& Landecker, T.~L.\ 1994, \aap, 291, 295 

\bibitem[Humphreys \& Davidson(1994)]{hum94} Humphreys, 
R.~M., \& Davidson, K.\ 1994, \pasp, 106, 1025 

\bibitem[Kn{\"o}dlseder(2000)]{kno00} Kn{\"o}dlseder, J.\ 
2000, \aap, 360, 539 

\bibitem[Kutner \& Ulich(1981)]{kut81} Kutner, M.~L., \& 
Ulich, B.~L.\ 1981, \apj, 250, 341 

\bibitem[Langer et al.(1994)]{lan94} Langer, N., Hamann, W.-R., 
Lennon, M., Najarro, F., Pauldrach, A.~W.~A., \& Puls, J.\ 1994, 
\aap, 290, 819 

\bibitem[Lequeux(2005)]{leq05} Lequeux, J.\ 2005, The 
interstellar medium. Astronomy and astrophysics library, 
Berlin: Springer, 2005

\bibitem[Maeder \& Meynet(1994)]{mae94} Maeder, A., \& Meynet, G. 1994, 
A\&A, 287, 803 

\bibitem[Marston et al.(1999)]{mar99} Marston, A.~P., Welzmiller, J., 
Bransford, M.~A., Black, J.~H., \& Bergman, P.\ 1999, \apj, 518, 769 

\bibitem[Massey \& Thompson(1991)]{mas91} Massey, P., \& 
Thompson, A.~B.\ 1991, \aj, 101, 1408 

\bibitem[Meaburn(2001)]{mea01} Meaburn, J.\ 2001, ASP 
Conf.~Ser.~233: P Cygni 2000: 400 Years of Progress, 233, 253 

\bibitem[Miller \& Chu(1993)]{mil93} Miller, G.~J., \& Chu, 
Y.-H.\ 1993, \apjs, 85, 137 

\bibitem[Moore et al.(2000)]{moo00} Moore, B.~D., Hester, 
J.~J., \& Scowen, P.~A.\ 2000, \aj, 119, 2991 

\bibitem[Nota \& Clampin(1997)]{not97} Nota, A., \& Clampin, M.\ 
1997, ASP Conf.~Ser.~120: Luminous Blue Variables: Massive Stars 
in Transition, 120, 303 

\bibitem[Nota et al.(2002)]{not02} Nota, A., Pasquali, A., 
Marston, A.~P., Lamers, H.~J.~G.~L.~M., Clampin, M., \& Schulte-Ladbeck, 
R.~E.\ 2002, \aj, 124, 2920 

\bibitem[O'Hara et al.(2003)]{oha03} O'Hara, T.~B., Meixner, 
M., Speck, A.~K., Ueta, T., \& Bobrowsky, M.\ 2003, \apj, 598, 1255 

\bibitem[Oka et al.(2001)]{oka01} Oka, T., et al.\ 2001, 
\apj, 558, 176 

\bibitem[Redman et al.(2003)]{red03} Redman, R.~O., Feldman, 
P.~A., Wyrowski, F., C{\^o}t{\'e}, S., Carey, S.~J., \& Egan, M.~P.\ 2003, 
\apj, 586, 1127

\bibitem[Rizzo et al.(2001a)]{riz01a} Rizzo, J.~R., 
Mart{\'{\i}}n-Pintado, J., \& Mangum, J.~G.\ 2001a, \aap, 366, 146 

\bibitem[Rizzo et al.(2001b)]{riz01b} Rizzo, J.~R., 
Mart{\'{\i}}n-Pintado, J., \& Henkel, C.\ 2001b, \apjl, 553, L181 

\bibitem[Rizzo et al.(2003a)]{riz03a} Rizzo, J.~R., 
Mart{\'{\i}}n-Pintado, J., \& Desmurs, J.-F.\ 2003a, IAU Symposium, 212, 740 

\bibitem[Rizzo et al.(2003b)]{riz03b} Rizzo, J.~R., 
Mart{\'{\i}}n-Pintado, J., \& Desmurs, J.-F.\ 2003b, IAU Symposium, 212, 742 

\bibitem[Rizzo et al.(2003c)]{riz03c} Rizzo, J.~R., 
Mart{\'{\i}}n-Pintado, J., \& Desmurs, J.-F.\ 2003c, \aap, 411, 465 

\bibitem[Schneider et al.(2006)]{sch06} Schneider, N., Bontemps, S., Simon, 
R., Jakob, H., Motte, F., Miller, M., Kramer, C., \& Stutzki, J.\ 2006, 
\aap, 458, 855 

\bibitem[Smith et al.(2003)]{smi03} Smith, N., Gehrz, R.~D., 
Hinz, P.~M., Hoffmann, W.~F., Hora, J.~L., Mamajek, E.~E., \& Meyer, M.~R.\ 
2003, \aj, 125, 1458 

\bibitem[Smith \& Owocki(2006)]{smi06} Smith, N., \& Owocki, 
S.~P.\ 2006, \apjl, 645, L45 

\bibitem[St-Louis et al.(1998)]{stl98} St-Louis, N., Doyon, 
R., Chagnon, F., \& Nadeau, D.\ 1998, \aj, 115, 2475

\bibitem[Trams et al.(1997)]{tra97} Trams, N.~R., Voors, 
R.~H.~M., \& Waters, L.~B.~F.~M.\ 1997, \apss, 255, 195 

\bibitem[Trams et al.(1999)]{tra99} Trams, N.~R., van Tuyll, 
C.~I., Voors, R.~H.~M., de Koter, A., Waters, L.~B.~F.~M., \& Morris, 
P.~W.\ 1999, LNP Vol.~523: IAU Colloq.~169: Variable and Non-spherical 
Stellar Winds in Luminous Hot Stars, 523, 71 

\bibitem[van Marle et al.(2004)]{mar04} van Marle, A.~J., Langer, 
N., \& Garc\'{\i}a-Segura, G.\ 2004, RMA\&A Conf.\ Ser., 22, 136

\bibitem[Voors et al.(2000)]{voo00} Voors, R.~H.~M., Geballe, T.~R., Waters, 
L.~B.~F.~M., Najarro, F., \& Lamers, H.~J.~G.~L.~M.\ 2000, \aap, 362, 236 

\bibitem[Waters et al.(1996)]{wat96} Waters, L.~B.~F.~M., Izumiura, H., Zaal, 
P.~A., Geballe, T.~R., Kester, D.~J.~M., \& Bontekoe, T.~R.\ 1996, \aap, 313, 866 

\bibitem[Wendker et al.(1991)]{wen91} Wendker, H.~J., Higgs, 
L.~A., \& Landecker, T.~L.\ 1991, \aap, 241, 551 

\bibitem[Wilson \& Matteucci(1992)]{wil92} Wilson, T.~L., \& 
Matteucci, F.\ 1992, \aapr, 4, 1 

\end{thebibliography}
\end{document}